\documentclass{article}

\usepackage[utf8]{inputenc}
\usepackage[margin=1in]{geometry}
\usepackage[affil-it]{authblk}
\usepackage{natbib}
\usepackage{bm}
\usepackage{amsmath,amsfonts,amssymb,amsthm}
\usepackage{newpxtext}
\usepackage[color=lightgray]{todonotes}
\usepackage[bf,font={small,sl}]{caption}
\usepackage{booktabs}

\usepackage{setspace}
\graphicspath{{figures/}}
\title{The Langevin diffusion as a continuous-time model of animal movement and habitat selection}
\author{Théo Michelot$^1$\footnote{Corresponding author: tmichelot1@sheffield.ac.uk}, Marie-Pierre Étienne$^2$, Pierre Gloaguen$^3$}
\affil{$^1$University of Sheffield , $^2$ AgroCampus Ouest, $^3$ AgroParisTech}
\date{}

\newcommand{\rmd}{\text{d}}
\newcommand{\Cov}{\mathbb{C}\text{ov}}
\newcommand{\ud}{\pi}
\newcommand{\td}[1]{q_{\Delta_{#1}}}
\newcommand{\X}{\bm{X}}

\newcommand{\hg}{\hat{\gamma}}
\newcommand{\hb}{\hat{\bm\beta}}
\newcommand{\hnu}{\hat{\bm{\nu}}}
\begin{document}
  \maketitle

\begin{abstract}
\noindent
1. The utilisation distribution describes the relative probability of use of a spatial unit by an animal. It is natural to think of it as the long-term consequence of the animal's short-term movement decisions: it is the accumulation of small displacements which, over time, gives rise to global patterns of space use. However, most utilisation distribution models either ignore the underlying movement, assuming the independence of observed locations, or are based on simplistic Brownian motion movement rules. \\ 
2. We introduce a new continuous-time model of animal movement, based on the Langevin diffusion. This stochastic process has an explicit stationary distribution, conceptually analogous to the idea of the utilisation distribution, and thus provides an intuitive framework to integrate movement and space use. We model the stationary (utilisation) distribution with a resource selection function to link the movement to spatial covariates, and allow inference into habitat selection. \\
3. Standard approximation techniques can be used to derive the pseudo-likelihood of the Langevin diffusion movement model, and to estimate habitat preference and movement parameters from tracking data. We investigate the performance of the method on simulated data, and discuss its sensitivity to the time scale of the sampling. We present an example of its application to tracking data of Stellar sea lions (\emph{Eumetopias jubatus}).\\
4. Due to its continuous-time formulation, this method can be applied to irregular telemetry data. It provides a rigorous framework to estimate long-term habitat selection from correlated movement data.  
\end{abstract}

\vspace{1em}
\noindent
{\bf Keywords:} animal movement, continuous time, resource selection, step selection, Langevin diffusion, potential function, utilisation distribution
 \vspace{1em}

\section{Introduction}
A crucial concept in animal ecology is the utilisation distribution, "the probability density function that gives the probability of finding an animal at a particular location" \citep{anderson1982home}. In recent decades, improvements in tracking technologies have produced large amounts of animal location data, at a high spatio-temporal resolution. Statistical methods have been developed to estimate the UD from telemetry observations, and to link animal movement to habitat preferences and space use \citep{hooten2017}. 
Most popular approaches to estimate the utilisation distribution from tracking data are non-parametric, and use empirical histograms \citep{nielson2013estimating} or kernel density estimation \citep{anderson1982home, worton1989, fleming2015}. More recently, interpolation methods involving Brownian bridges have also been used \citep{horne2007,kranstauber2012dynamic}. 
 A major limitation of such methods, however, is that the estimation of the utilisation distribution is disconnected from the movement itself, as they either ignore the sequential structure of data \citep{anderson1982home, worton1989, nielson2013estimating}, or make unrealistic Brownian assumptions about the animal's movement \citep{horne2007, kranstauber2012dynamic, fleming2015}.

In those approaches, a (generally two-dimensional) density function is estimated, and can be related to environmental covariates using regression techniques \citep{millspaugh2006analysis, long2009comparison, nielson2013estimating,  zhang2014ecological}. 
The regression function, formulated in terms of spatial covariates of interest, is often defined as a resource selection function (RSF, \citealp{manly2007}). 
A RSF links the distribution of observed locations of an animal to the distribution of resources (or other spatial covariates), to infer habitat characteristics that are preferred (or ``selected'') by the animal. 
It is based on the idea that, knowing the habitat composition of a spatial unit, we can predict its long-term utilisation.
 However, resource selection models usually assume that telemetry observations are independent, which is unrealistic for high-frequency movement data.

It is natural to think of the utilisation distribution as a consequence of the movement, which itself depends on the environment, such that short-term movement decisions give rise to long-term space use. 
This idea motivates the development of more mechanistic approaches that link the animal's movement to its environment, and, ultimately, a mechanistic movement model with an explicit steady-state distribution, representing the utilisation distribution.

Following this idea, step selection functions model the likelihood of a step between two points in space as a combination of a movement kernel and a habitat selection function \citep{fortin2005, forester2009, thurfjell2014applications}. 
The parameters of a step selection function describe preference at a local (step-by-step) level, and strongly depend on the choice of the movement kernel. 
As such, it is unclear how they can be used to infer global space use. 
\cite{potts2014} and \cite{avgar2016} have suggested numerical methods to approximate the utilisation distribution underlying a step selection function model, but it does not take a parametric closed form.

\cite{hanks2015continuous} proposed a continuous-time discrete-space model to link movement to environmental drivers. 
In their framework, the movement is considered as a continuous-time Markov process on a discrete grid of spatial cells. The spatial grid is usually chosen as the grid on which the spatial covariates are measured, and the observed locations are binned in the cells. 
\cite{wilson2018estimating} argued that the limiting distribution of that movement model can be interpreted as the utilisation distribution of the animal, and proposed a method to estimate it on a discrete grid. A drawback of that approach is that it describes movement on a discrete spatial grid, and its formulation is therefore tied to a particular space discretization.

Recently, \cite{michelot2017} proposed a step selection model, formulated in terms of an explicit utilisation distribution. 
Their approach describes individual movement as a Markov chain in continuous space, whose stationary distribution is the utilisation distribution. 
In particular, they suggest that Markov chain Monte Carlo (MCMC) algorithms, which are used to construct Markov chains with a given stationary distribution, can be viewed as movement models.

Others have described the position of an animal as a diffusion process which follows the gradient of a potential surface \citep{brillinger2010, preisler2013, gloaguen2018}. The surface measures the potential interest for the animal, but it is not directly connected to the utilisation distribution. These approaches offer a wide variety of flexible models to describe movement, but their link to the utilisation distribution is unclear in the existing literature. Indeed, potential-based models are often based on diffusion processes that are not stationary \citep{gloaguen2018}, or lead to unrealistically simple utilisation distributions. For example, the stationary distribution is uniform over the study region for Brownian motion movement models \citep{skellam1951random}, and it is a normal distribution for Ornstein-Uhlenbeck based movement models \citep{blackwell1997random}.

In this work, we describe a new continuous-time and space mechanistic movement model. 
The animal's position is modelled as a diffusion process with a drift towards the gradient of its stationary (utilisation) distribution, bringing together the ideas of \cite{brillinger2010} and \cite{michelot2017}.
 As in \cite{wilson2018estimating}, the limiting distribution of the process is the utilisation distribution.
The movement model is based on the Langevin diffusion, which has also been used to construct a MCMC algorithm \citep{roberts1998}. 
As this model belongs to the class of potential-based models, inference can be performed from movement data using different estimation methods for stochastic differential equations (SDEs), such as pseudo-likelihood methods which are simple to implement \citep{gloaguen2018}.
We show here how this parametric model can also be linked to step selection approaches when the utilisation distribution is parameterized as a simple function of environmental covariates.
Point estimators and confidence intervals of habitat selection parameters can easily be derived in a classical approximated inference framework.

In Section \ref{sec:model}, the proposed movement model is formulated in its general form, and in a specific "covariate-dependent" form, using a resource selection function. 
Section \ref{sec:inference} describes a pseudo-likelihood method based on the Euler discretization scheme, for the estimation of habitat selection parameters from telemetry data. 
In Section \ref{sec:sim}, we assess the performance of inference methods in simulations, and we discuss conditions under which the model parameters can be recovered.
In Section \ref{sec:casestudy}, we present the analysis of three trajectories of Stellar Sea Lions (\emph{Eumetopias jubatus}) with the Langevin movement model, with four environmental covariates as potential drivers of their movement.

\section{Langevin movement model}
\label{sec:model}
\subsection{General formulation}
We denote by $\X_t \in \mathbb{R}^d$ the location of an animal in $d$-dimensional space at time $t \geq 0$, and $\ud : \mathbb{R}^d \rightarrow \mathbb{R}$ its utilisation distribution \citep{worton1989}. The utilisation distribution is the probability density function such that, for any area $A \subset \mathbb{R}^d$,
\begin{equation}
\label{eq:def_UD}
\mathbb{P}(\bm{X}_t \in A) = \int_A \pi(\bm{z}) \rmd \bm{z}
\end{equation}
 In the following, we will focus on the case $d=2$, by far the most common in movement ecology, although the framework could be extended to higher dimensions.

We propose to describe the continuous-time location process of the animal $(\X_t)_{t \geq 0}$  with a Langevin diffusion for the density $\ud$, defined as the solution to the stochastic differential equation
\begin{equation}
	\label{eq:langevin_basic}
	\rmd\X_t = \dfrac{1}{2} \nabla \log \ud(\X_t) \rmd t + \rmd \bm{W}_t,
\end{equation}
where $\bm{W}_t$ is a standard Brownian motion, $\nabla$ is the gradient operator, and with initial condition $\X_0 =\bm{x}_0$. 
Under some easily-satisfied technical conditions \citep[that can be found in][]{dalalyan2017}, Equation \eqref{eq:langevin_basic} has a unique solution, which is a continuous-time continuous Markov process. 
It describes the animal's movements as the combination of a drift towards higher values of its utilisation distribution $\ud$ (informed by the gradient of $\log \ud$), and a random component given by the Brownian motion. 
Crucially, the solution is a continuous-time Markov process whose asymptotic stationary distribution is $\ud$ as defined in Equation \eqref{eq:def_UD} \citep{roberts1996}. 
The Langevin diffusion is thus a natural choice for the basis of a continuous-time model of animal movement, with a known steady-state distribution.

In its standard formulation, however, the Langevin diffusion cannot readily be used to model animal movement. 
Indeed, the speed of the process described above is only determined by the shape of the underlying utilisation distribution $\pi$, whereas it should be possible for two animals to move at different speeds on the same long-term distribution of space use. 
To allow for this flexibility, we introduce an additional parameter $\gamma^2$, and we define the Langevin movement model (with speed), as the solution to
\begin{equation}
	\label{eq:langevin}
	\rmd\X_t = \dfrac{\gamma^2}{2} \nabla \log \ud(\X_t) \rmd t + \gamma \rmd \bm{W}_t,\quad \X_0 =\bm{x}_0.
\end{equation}

\begin{figure}[htbp]
\centering
\begin{tabular}{cc}
\includegraphics[width = 0.48\textwidth]{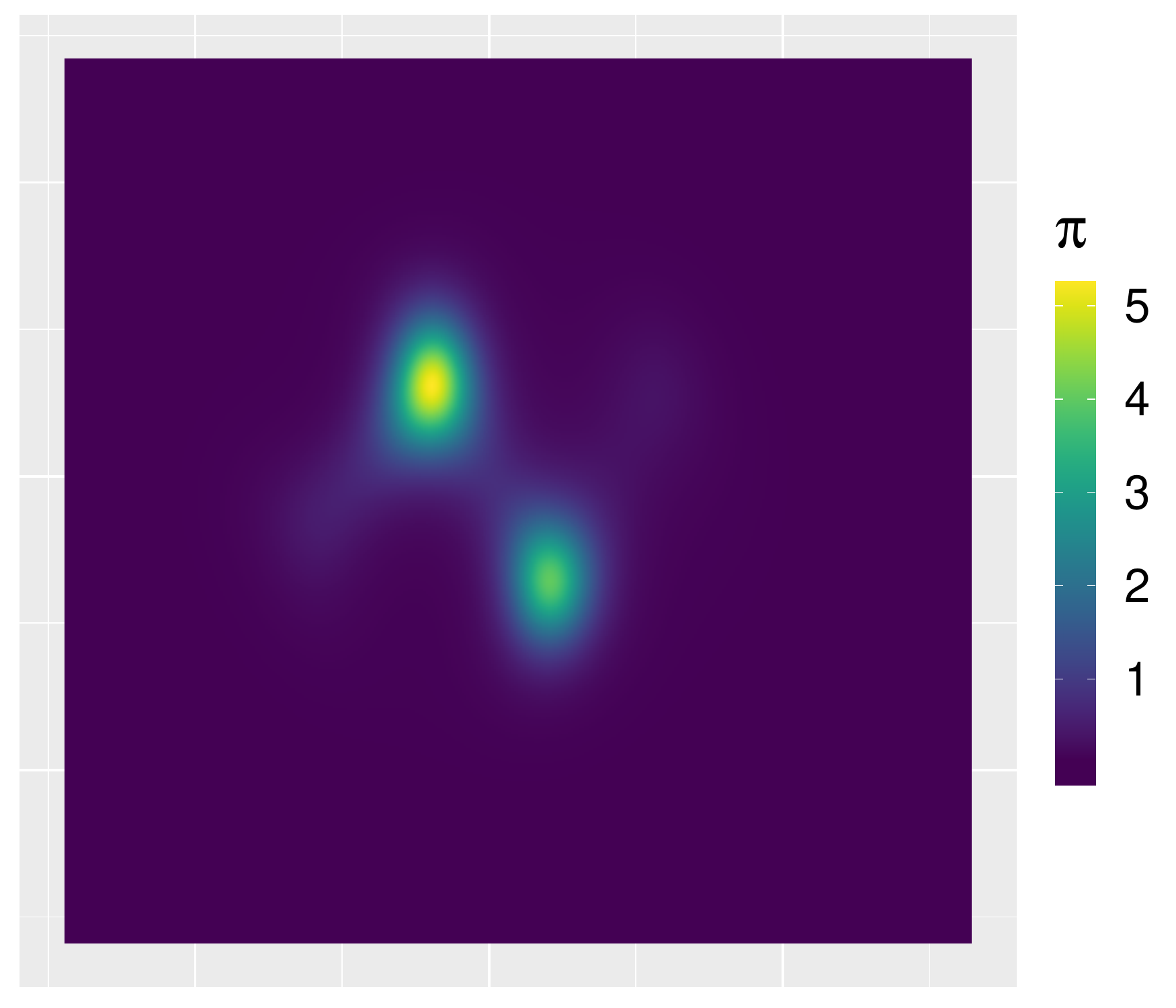}&
\includegraphics[width = 0.48\textwidth]{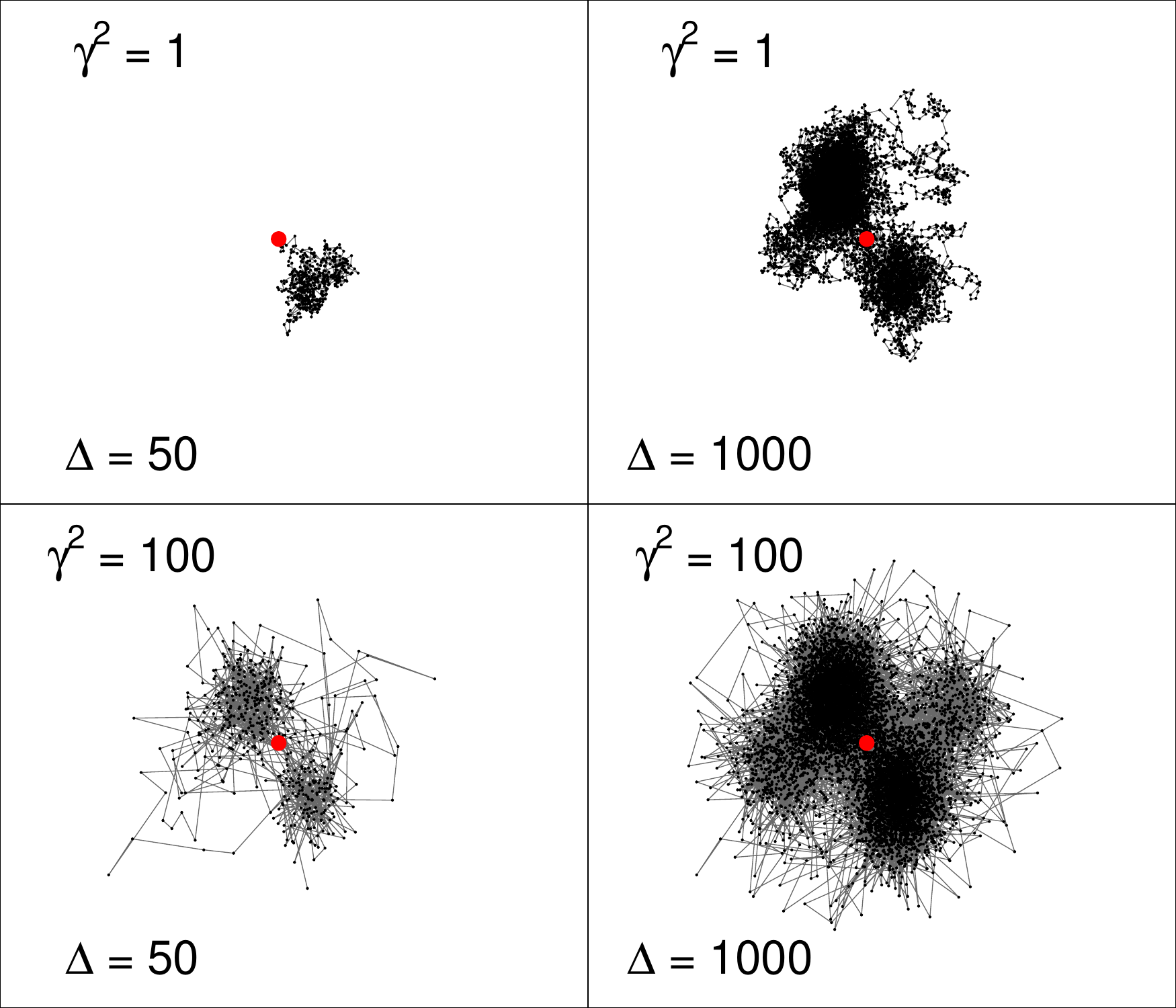}
\end{tabular}
\caption{\label{fig:AnalyticUD} \textit{Left:} Artificial utilisation distribution $\pi$. \textit{Right:} Trajectories simulated from the Langevin movement model on $\pi$, with two different values of the speed parameter $\gamma^2$ (1 and 100), after $\Delta=50$ and $\Delta=1000$ time units. Although the process with $\gamma^2=1$ is much slower to explore space, the properties of the Langevin equation guarantee that both processes have the same stationary distribution $\pi$.}
\end{figure}

In the following, $\gamma^2$ will be referred to as the speed parameter. 
The model is parameterized in terms of $\gamma^2$ (rather than $\gamma$) as it is also the variance parameter of the random Brownian motion component. 
Figure \ref{fig:AnalyticUD} shows two tracks simulated from the Langevin movement model on an artificial utilisation distribution, for two different values of $\gamma^2$.
 Although the two tracks explore space at very different speeds, they have the same equilibrium distribution.

\subsection{Including covariates}
\label{sec:covariates}

We link the utilisation distribution of the animal to spatial covariates with the standard parametric form of resource selection functions (RSF),
\begin{equation}
	\label{eq:rsf}
	\ud(\bm{x} \vert \bm\beta) = \dfrac{\exp\left(\sum_{j=1}^J \beta_j c_j (\bm{x}) \right)}
    {\int_\Omega \exp\left(\sum_{j=1}^J \beta_j c_j (\bm{z}) \right) d\bm{z}} ,
\end{equation}
where $c_j(\bm{x})$ is the value of the $j$-th covariate at location $\bm{x}$, $\Omega \subset \mathbb{R}^d$ is the study region, and $\bm\beta = (\beta_1,\dots,\beta_J)'$ is a vector of unknown parameters. 
Larger values of $\beta_j$ indicate stronger selection of the $j$-th covariate. 
The denominator in the right-hand side of Equation \eqref{eq:rsf} is a normalising constant, and is necessary to ensure that $\ud(\bm{x} \vert \bm\beta)$ is a probability density function with respect to $\bm{x}$.

Note that Equation \eqref{eq:langevin} requires $\log \ud$ to be a smooth function, i.e.\ with continuous first-order partial derivatives. 
If $\ud$ is modelled by a resource selection function (Equation \eqref{eq:rsf}), then
\begin{equation}
	\label{eq:gradlogpi}
    \nabla \log \ud (\bm{x} \vert \bm\beta) = \sum_{j=1}^J \beta_j \nabla c_j (\bm{x}).
\end{equation}

Therefore, it is supposed here that all covariates $c_j$ are differentiable, and that their gradient can be computed at each point $\bm{x}$, either analytically, or by numerical approximation. 
In most real data sets, the covariate functions $c_j$ are  measured at discrete points in space. 
There is generally no analytical form for the gradient, and it is necessary to interpolate the covariate fields so that its gradient can be approximated. 
In Sections \ref{sec:sim2} and \ref{sec:casestudy}, bilinear interpolation is considered  to obtain continuous covariate functions. 
As a consequence of this interpolation, the Langevin movement model cannot handle categorical covariates. 
Indeed, a categorical covariate field cannot be interpolated into a continuous function, and it is generally not possible to derive a measure of its gradient.

\section{Inference}
\label{sec:inference}

The continuous-time location process of the animal $(\X_t)_{t\geq 0}$ is observed discretely at times $t_0 < t_1< \dots < t_n$, and  these observations are denoted by $(\bm{x}_0, \bm{x}_1, \dots, \bm{x}_n)$. We consider
$J$ spatial covariates $c_1, \dots, c_J$, measured on a grid over the study region. 
$\bm\theta$ denotes the vector of all parameters of the Langevin movement model defined in Section \ref{sec:model}, i.e.\ $\bm\theta = (\beta_1,\dots,\beta_J,\gamma^2)$. 
This section describes an inference method to estimate $\bm\theta$, from telemetry and habitat data.

\subsection{Euler approximation of the likelihood}
\label{sec:Euler}

The likelihood of the observed locations, given $\bm\theta$, can be expressed using the \textit{transition density} of the process $(\X_t)_{t\geq 0}$. 
The transition density is the probability density function of the random variable $\bm{X}_{t + \Delta}$ given that $\bm{X}_t = \bm{x}_t$, and we denote it by $\td{}(\bm{x} \vert \bm{x}_t, \bm\theta)$. 
By the Markov property of the solution to Equation \eqref{eq:langevin}, and assuming that the first position is deterministic, the likelihood function is
\begin{equation}
\label{eq:likelihood}
L(\bm\theta; \bm{x}_{0:n}) = \prod_{i = 0}^{n - 1} \td{i}(\bm{x}_{i+1} \vert \bm{x}_i, \bm\theta),
\end{equation}
where $\bm{x}_{0:n}$ is shorthand for the set of observations, and $\Delta_i = t_{i+1} - t_i$.

As discussed in \cite{gloaguen2018}, in many practical cases, the density $\td{}$ is intractable, and the likelihood $L(\bm\theta; \bm{x}_{0:n})$ cannot be evaluated.
To circumvent this problem, pseudo-likelihood approaches can be used as approximations. 
In these approaches, the intractable transition density in Equation \eqref{eq:likelihood} is replaced by the p.d.f.\ of a known distribution (usually, Gaussian), with moments given by a discretisation scheme.

The most common pseudo likelihood approach for discretely observed diffusion is the Euler discretization scheme \citep{iacus2009}. 
In the Euler discretization (for $d = 2$), the transition density of the Langevin diffusion is approximated by the following Gaussian density between $t_i$ and $t_{i+1}$, for $i = 0, \dots, n-1$,
\begin{equation}
\label{eq:eulerApprox}
	\X_{i+1} \vert \lbrace \bm{X}_i = \bm{x}_i \rbrace = 
    	\bm{x}_i + \dfrac{\gamma^2 \Delta_i}{2} \nabla \log \pi(\bm{x}_i \vert \bm\beta) + \varepsilon_{i+1},\quad 
    	\varepsilon_{i+1} \overset{ind}{\sim} N \left( \bm{0} ,\ \gamma^2 \Delta_i \bm{I}_2 \right),
\end{equation}
where $\bm{I}_2$ is the $2 \times 2$ identity matrix.
 Under this approximation, the transition density of the process can then be written
\begin{equation*}
	q_{\Delta_i}(\bm{x}_{i+1} \vert \bm{x_i}, \bm\theta) = 
    \phi \left( \bm{x}_{i+1} \bigg\vert \bm{x}_i + 
    \dfrac{\gamma^2 \Delta_i}{2} \nabla \log \pi(\bm{x}_i \vert \bm{\beta});\ 
    \gamma^2 \Delta_i \bm{I}_2 \right),
\end{equation*}
where $\phi(\cdot \vert \bm\mu;\ \bm\Sigma)$ is the p.d.f.\ of the multivariate normal distribution with mean $\bm\mu$ and covariance matrix $\bm\Sigma$. This expression can be plugged into Equation \eqref{eq:likelihood} to obtain the likelihood of a track $\bm{x}_{0:n}$.

The Euler discretization can also be used to simulate (approximately) from the Langevin movement model, as illustrated in the simulations of Section \ref{sec:sim}. The quality of the scheme decreases for longer time steps of simulation (\citealp{kessler2012statistical}, Chapter 1).

\subsection{Maximum likelihood estimation}
\label{sec:MLE}

The pseudo-likelihood function could be optimised numerically to obtain estimates of all model parameters. However, if $\ud$ is modelled with the resource selection function of Equation \eqref{eq:rsf}, the maximum likelihood estimate~$\hat{\bm{\theta}}$ can simply be obtained using standard linear model equations. 

Plugging Equation \eqref{eq:gradlogpi} into Equation \eqref{eq:eulerApprox}, we can write a standard linear model, in the following matrix form. Let $\bm{Y}_i = (\bm{X}_{i+1} - \bm{X}_i)/\sqrt{\Delta_i}$ be the (two-dimensional) normalized random increment of the process between $t_i$ and $t_{i+1}$, and denote
\begin{equation*}
\bm{Y} =
\begin{pmatrix} 
	Y_{0,1} \\ 
	\vdots \\
    Y_{n - 1,1}\\ 
	Y_{0,2}\\ 
	\vdots  \\
    Y_{n - 1,2}\\ 
\end{pmatrix},
\qquad 
\bm{D} = \dfrac{1}{2} 
\begin{pmatrix}
	\frac{\partial c_1(\bm{x}_0)}{\partial z_1} & \frac{\partial c_2(\bm{x}_0)}{\partial z_1} & \ldots & 
    \frac{\partial c_J(\bm{x}_0)}{\partial z_1}\\
	\vdots & &  &\vdots \\
	\frac{\partial c_1(\bm{x}_{n-1})}{\partial z_1} & \frac{\partial c_2(\bm{x}_{n-1})}{\partial z_1} & \ldots & 
    \frac{\partial c_J(\bm{x}_{n-1})}{\partial z_1}\\
	\frac{\partial c_1(\bm{x}_0)}{\partial z_2} & \frac{\partial c_2(\bm{x}_0)}{\partial z_2} & \ldots & 
	\frac{\partial c_J(\bm{x}_0)}{\partial z_2}\\
	\vdots & &  &\vdots \\
	\frac{\partial c_1(\bm{x}_{n-1})}{\partial z_2} & \frac{\partial c_2(\bm{x}_{n-1})}{\partial z_2} & \ldots & 
    \frac{\partial c_J(\bm{x}_{n-1})}{\partial z_2}, \\
\end{pmatrix},
\end{equation*}
where $\bm{Y}_i = (Y_{i,1}, Y_{i,2})$, and $\partial/\partial z_i$ denotes the partial derivative with respect to the $i$-th spatial coordinate. 

Moreover, let $\bm{T}_{\Delta}$ be the $(2n) \times (2n)$ diagonal matrix with $i$-th and $(n+i)$-th diagonal terms equal to $\sqrt{\Delta_{i-1}}$, for $i = 1,\dots, n$. 
Then, the Euler approximation of the Langevin movement model can be rewritten as 
\begin{equation}
\label{eq:lin_model}
	\bm{Y} = (\bm{T}_{\Delta} \bm{D}) \bm\nu + \bm{E},
\end{equation}
where $\bm{E}$ is a $2n$-vector of $N(0, \gamma^2)$ variables, and $\bm\nu = \gamma^2 \bm\beta$. The estimators for $\bm\nu$ and $\gamma^2$ are derived from standard linear model theory, as
\begin{equation*}
	\hnu = \left (\bm{D}^{\intercal} \bm{T}_{\Delta}^2  \bm{D} \right)^{-1} 
    	\bm{D}^{\intercal} \bm{T}_{\Delta} \bm{Y},
\end{equation*}
and 
\begin{equation*}
	\hg^2 = \dfrac{1}{2n-J} \lVert \bm{Y} - \hat{\bm{Y}} \rVert^2,
\end{equation*}
where $\hat{\bm{Y}} = (\bm{T}_\Delta \bm{D}) \hat{\bm\nu}$ is the predicted value of $\bm{Y}$. 
In Appendix A, we show that an unbiased estimator of $\bm\beta$ is then given by  
$$\hb = \frac{(2n - J - 2)\hnu}{(2n - J)\hg^2},$$
with the following covariance formula
\begin{equation*}
	\Cov(\hat\beta_j, \hat\beta_k) = 
    	\frac{2\beta_j\beta_k}{2n - J - 4} + \frac{\Upsilon_{jk}}{\gamma^2} \left(1 + \frac{2}{2n - J - 4} \right),
\end{equation*}
where $\Upsilon_{jk} := \left[ (\bm{D}^{\intercal} \bm{T}_{\Delta}^2  \bm{D} )^{-1} \right]_{jk}$. Using the asymptotic normality of maximum likelihood estimators, we can obtain a confidence interval of any level $\alpha \in (0,0.5)$, that is valid for a large enough $n$. For each covariate coefficient $\beta_j$,

\begin{equation*}
	CI_\alpha(\beta_j) = \left[ \hat{\beta_j} + z_{\alpha / 2} \times \sqrt{\mathbb{V}(\hat\beta_j)} ; 
    ~~\hat{\beta_j} - z_{\alpha / 2} \times \sqrt{\mathbb{V}(\hat\beta_j)} \right]
\end{equation*}
where $z_{\alpha / 2}$ is the quantile of level $\alpha /2$ of a standard Gaussian distribution, and $\mathbb{V}(\hat\beta_j) = \Cov(\hat\beta_j, \hat\beta_j)$. The detail of the derivation is given in Appendix \ref{append:beta}.

The results above derive from the use of a linear model to approximate the solution to a SDE. 

For the Langevin movement model based on a RSF, as defined in Section \ref{sec:model}, the Euler approximation therefore provides explicit estimates and confidence intervals. 
Note that the Euler estimator is \textit{biased} due to the approximation made in Equation \eqref{eq:eulerApprox} (see \citealp{kessler2012statistical}, Chapter 1). 
Therefore, both the estimate and the confidence interval must be interpreted with caution, as they depend on the quality of the scheme. 
The potential use of other discretization schemes is discussed in Section \ref{sec:Discussion}.

\section{Simulation study}
\label{sec:sim}

In this section, we assess the performance of the inference method described in Section \ref{sec:inference} in two simulation scenarios.
In both cases, we simulate movement tracks from the Langevin process, using the Euler discretization given in Equation \ref{eq:eulerApprox}. We simulate covariates and define an artificial utilisation distribution, expressed as a resource selection function, as shown in Equation \eqref{eq:rsf}.
The objective is to recover the habitat selection parameters $\{ \beta_1, \dots, \beta_J \}$ and the speed parameter $\gamma^2$.

\subsection{Scenario 1}
\label{sec:sim1}
We first consider a fully controlled simulation scenario, where the covariate fields are given by smooth analytical functions. 
In this idealized case, the gradient of the covariate functions, and thus of the utilisation distribution, can be calculated exactly at any point in the plane. 
The utilisation distribution $\pi$ is defined as a RSF (Equation \eqref{eq:rsf}) of three covariates $c_1$, $c_2$ and $c_3$, given by
\begin{align*}
	c_j(\bm{z}) & = \alpha_j\exp(-(\bm{z} - \bm{a}^j)^\intercal \bm{\Sigma}^j (\bm{z} - \bm{a}^j)) 
    	\times \sin \left( \omega_1^j (z_1 - a_1^j) \right) 
    	\times \sin \left( \omega_2^j (z_1 - a_2^j) \right),~ j = 1, 2,\\ 
	c_3(\bm{z}) & = \parallel \bm{z} \parallel^2,
\end{align*}
where $alpha_j, \bm{a}^j = (a_1^j, a_2^j),~ \bm{\omega}^j = (\omega_1^j, \omega_2^j)$, and $\bm{\Sigma}^j = \text{diag} \lbrace \sigma_1^j, \sigma_2^j \rbrace$ are known simulation parameters whose values are given in Appendix \ref{append:simpar}. 
For the simulations, we choose the resource selection parameters $\beta_1 = -1$, $\beta_2 = 0.5$, and $\beta_3 = -0.05$, and the speed parameter $\gamma^2=1$.

\begin{figure}[htbp]
	\centering
	\includegraphics[width = 0.32\textwidth]{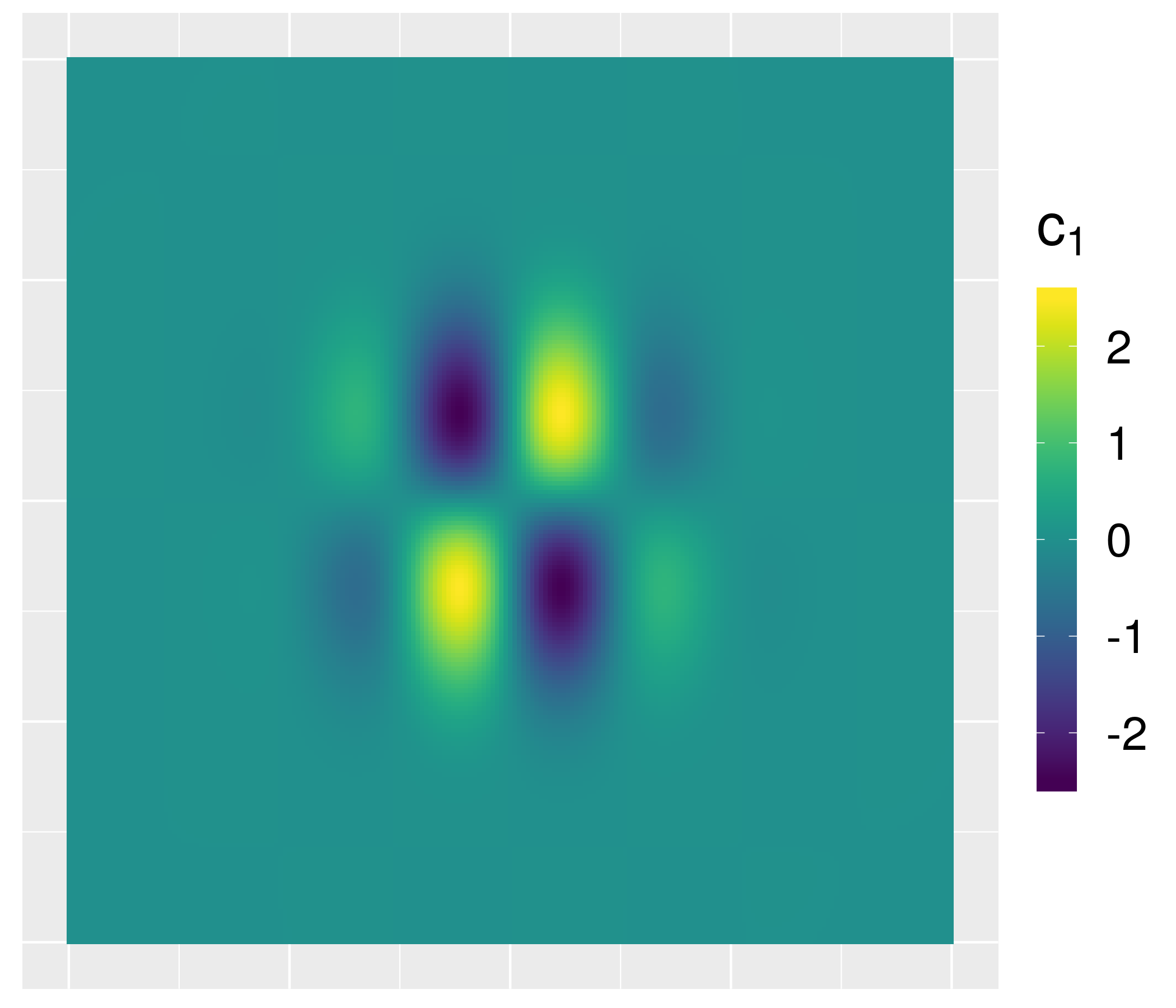}
    \includegraphics[width = 0.32\textwidth]{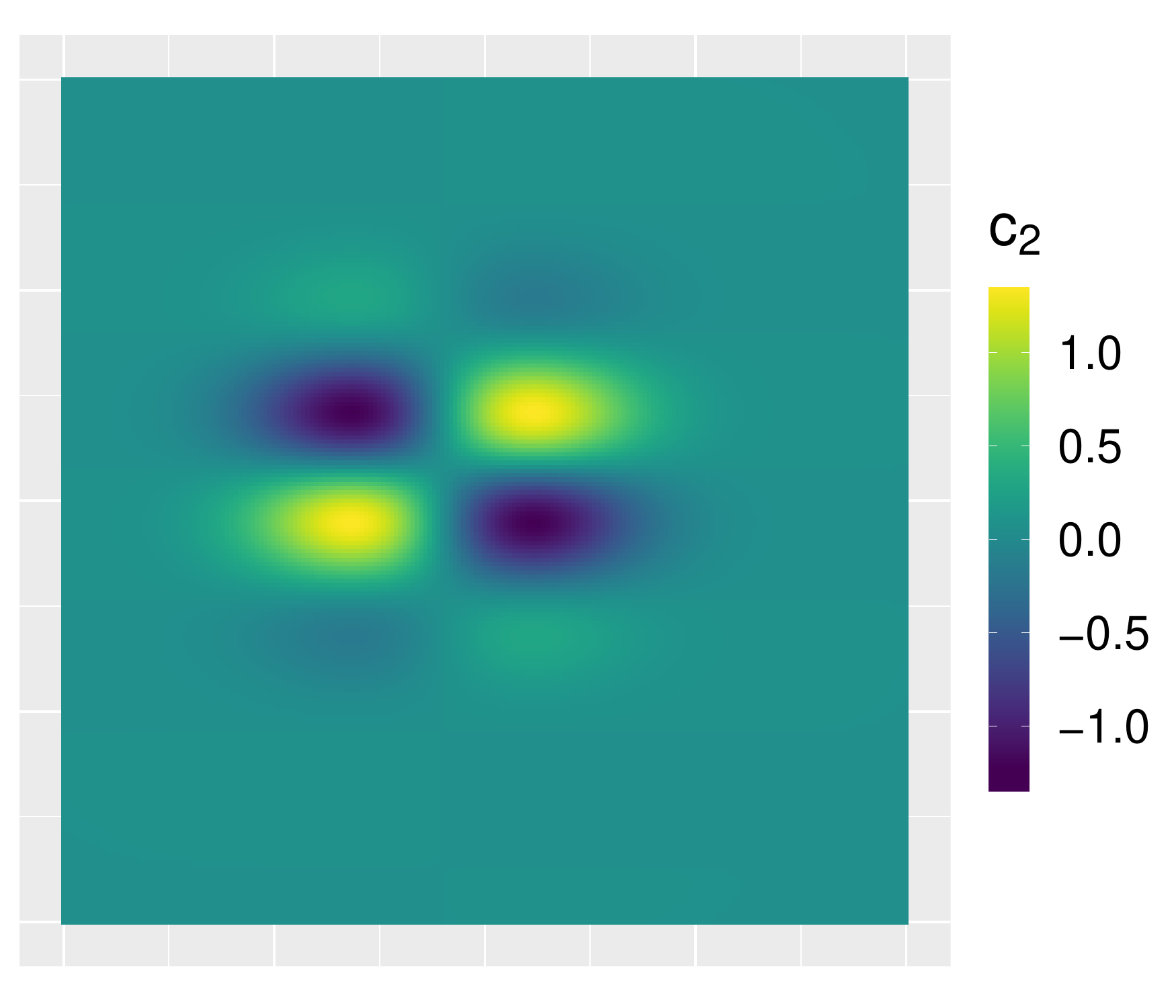}
    \includegraphics[width = 0.32\textwidth]{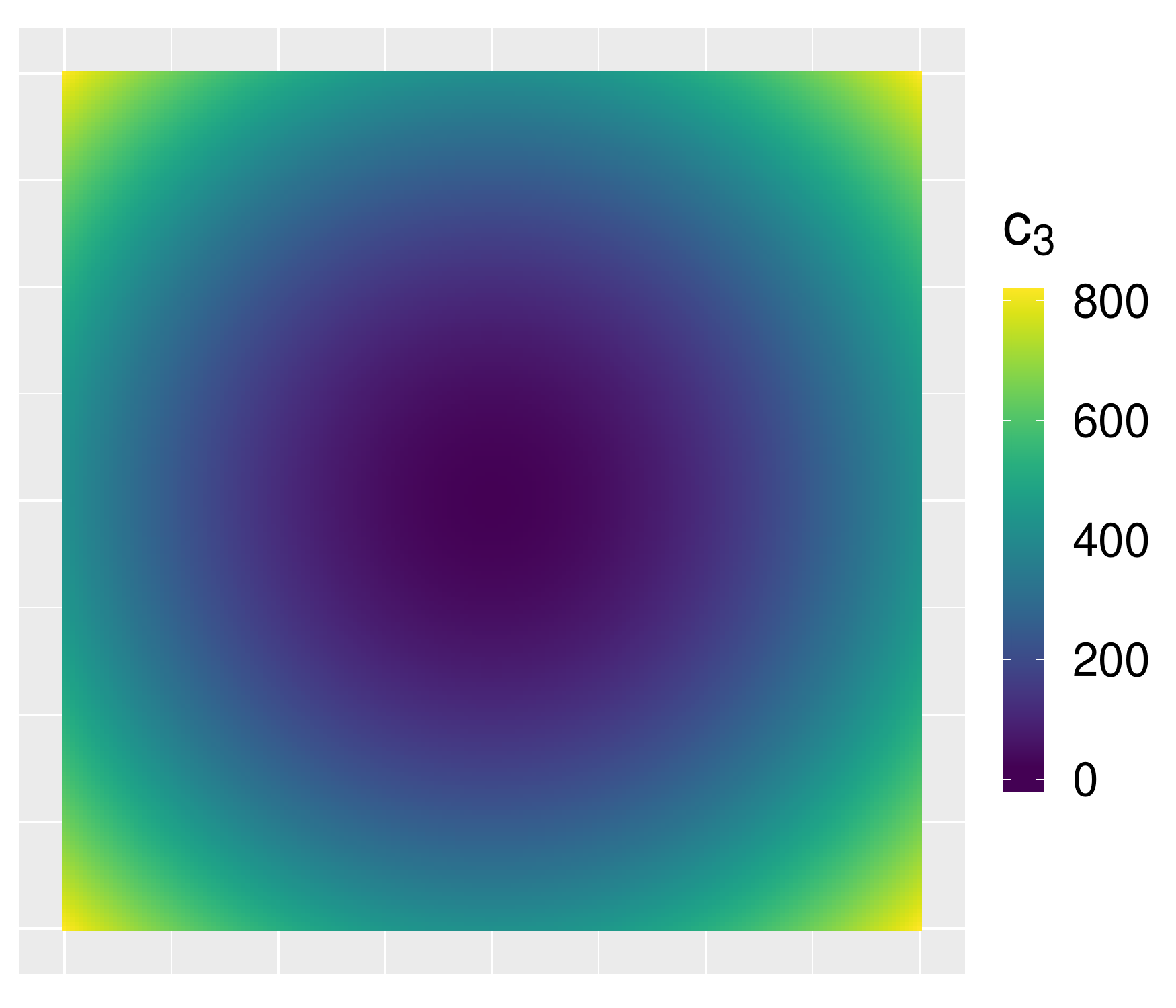}
	\caption{\label{fig:AnalyticCov} Artificial covariates fields for the simulation scenario of Section \ref{sec:sim1}.}
\end{figure}

The first two covariates are smooth functions, for which the gradient can easily be derived. 
The third covariate is the squared distance to the centre of the map, and is used to include a weak force of attraction towards the centre (here, the point $(0,0)$, somewhat related to the home range of the individual). 
These three covariates functions are shown in Figure \ref{fig:AnalyticCov}. 

Inference was performed independently on 600 data sets. Each data set was a trajectory of 300 points, simulated from the Langevin movement model. The tracks were first generated at a fine time resolution ($\Delta = 0.01$), to minimise the effect of the Euler approximation, and they were then thinned to time intervals of $0.5$ time units.

We estimated all model parameters using the Euler method, presented in Section \ref{sec:MLE}. We considered two different settings: (i) the true analytic gradient is used in the estimation, and (ii) the covariates are discretized on a $8 \times 8$ regular grid, and the gradient is obtained through the interpolation of the covariates.
This second setting corresponds to the more realistic case where covariates are only observed on a discrete grid, and the gradient needs to be approximated. 
The gradient approximations were performed for the covariates $c_1$ and $c_2$ using the R package nloptr \citep{ypma2014nloptr}. The gradient of the Euclidean distance $c_3$ is computed exactly in both cases, as it would be in a real analysis.

\begin{figure}[htbp]
\centering
\includegraphics[width = 0.6\textwidth]{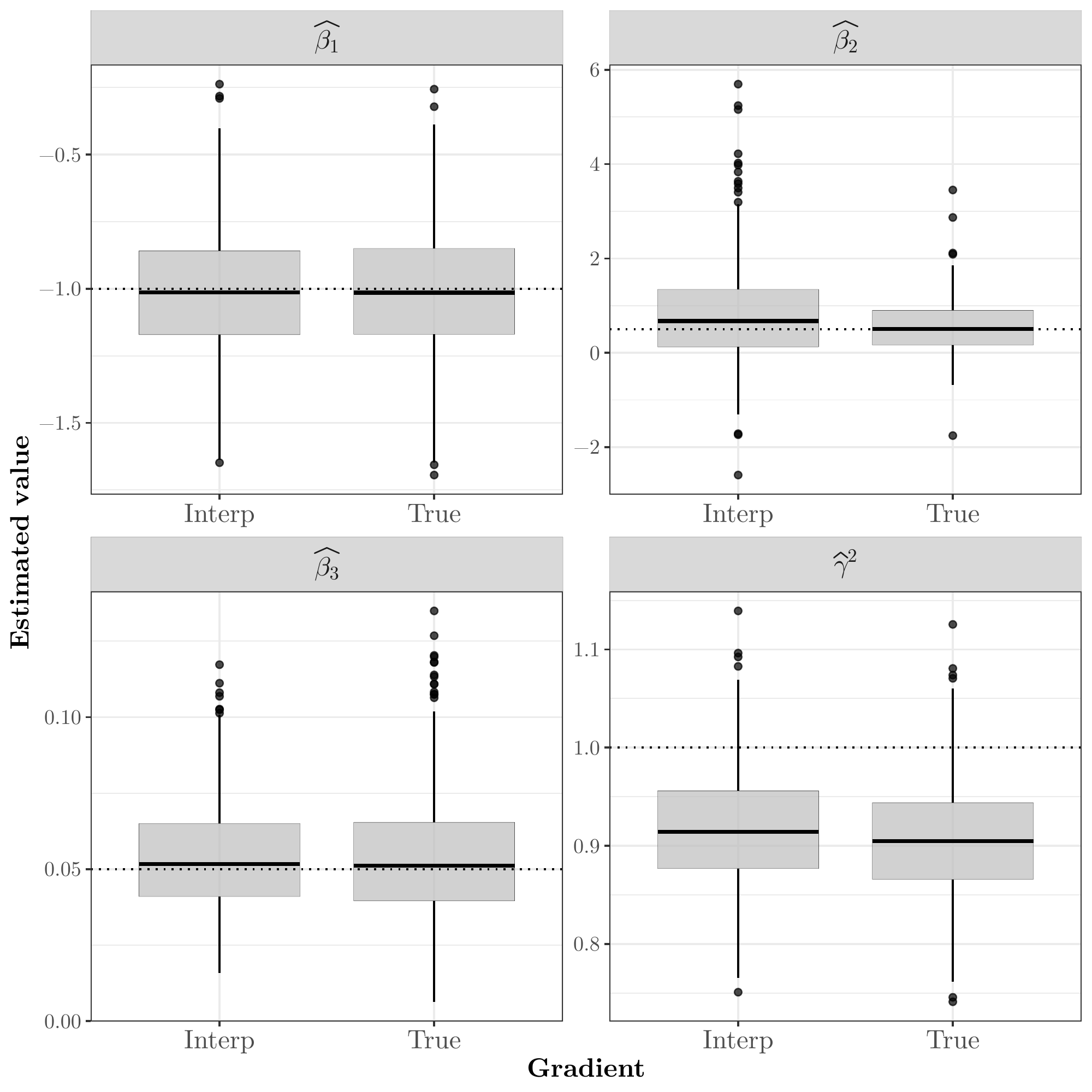}
\caption{\label{fig:EstSc1} Estimates for model parameters on 600 experiments replications of scenario 1. The dotted lines show the real values used in the simulations.}
\end{figure}

Boxplots of the parameter estimates in the 600 replications are shown in Figure \ref{fig:EstSc1}. All parameters were correctly estimated in this benchmark scenario, even when the covariates were discretized to a coarse grid.
    
\subsection{Scenario 2}
\label{sec:sim2}
We considered a second simulation scenario, with randomly-generated covariate fields on a discrete grid, more similar to real environmental data. The main objective of this scenario is to investigate the effect of the sampling frequency on the estimation.

\begin{figure}[htbp]
	\centering
	\includegraphics[width=0.32\textwidth]{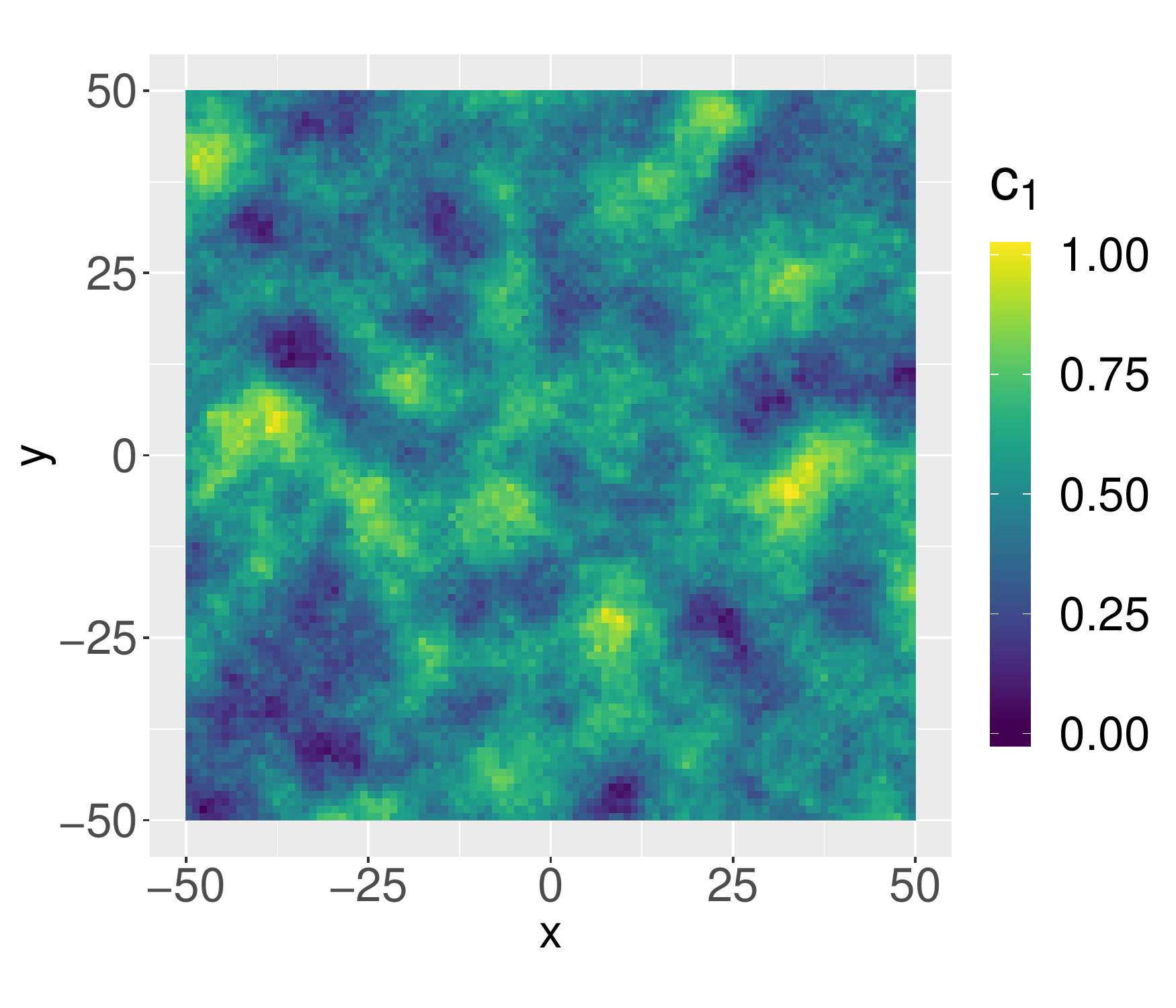}
    \includegraphics[width=0.32\textwidth]{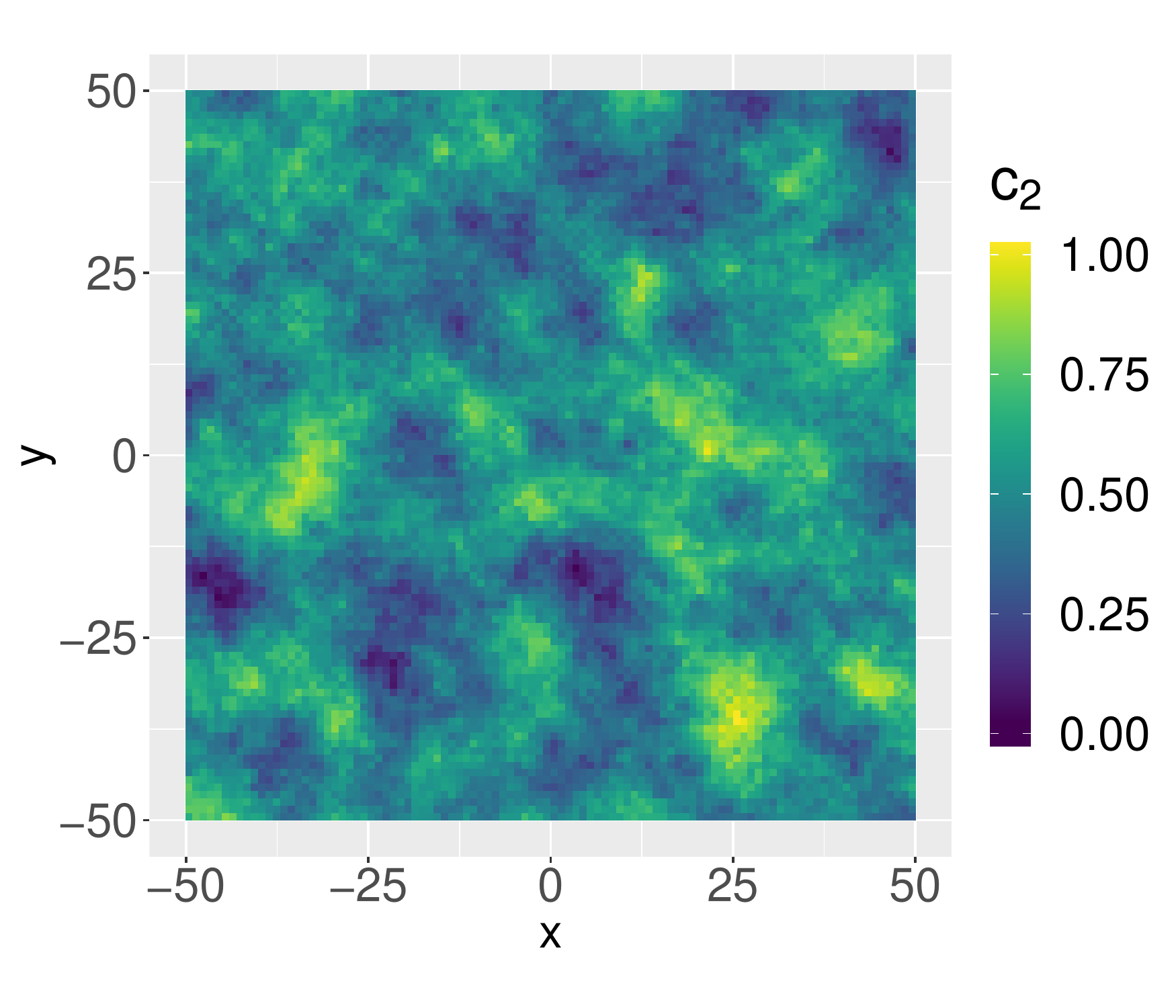}
    \includegraphics[width=0.32\textwidth]{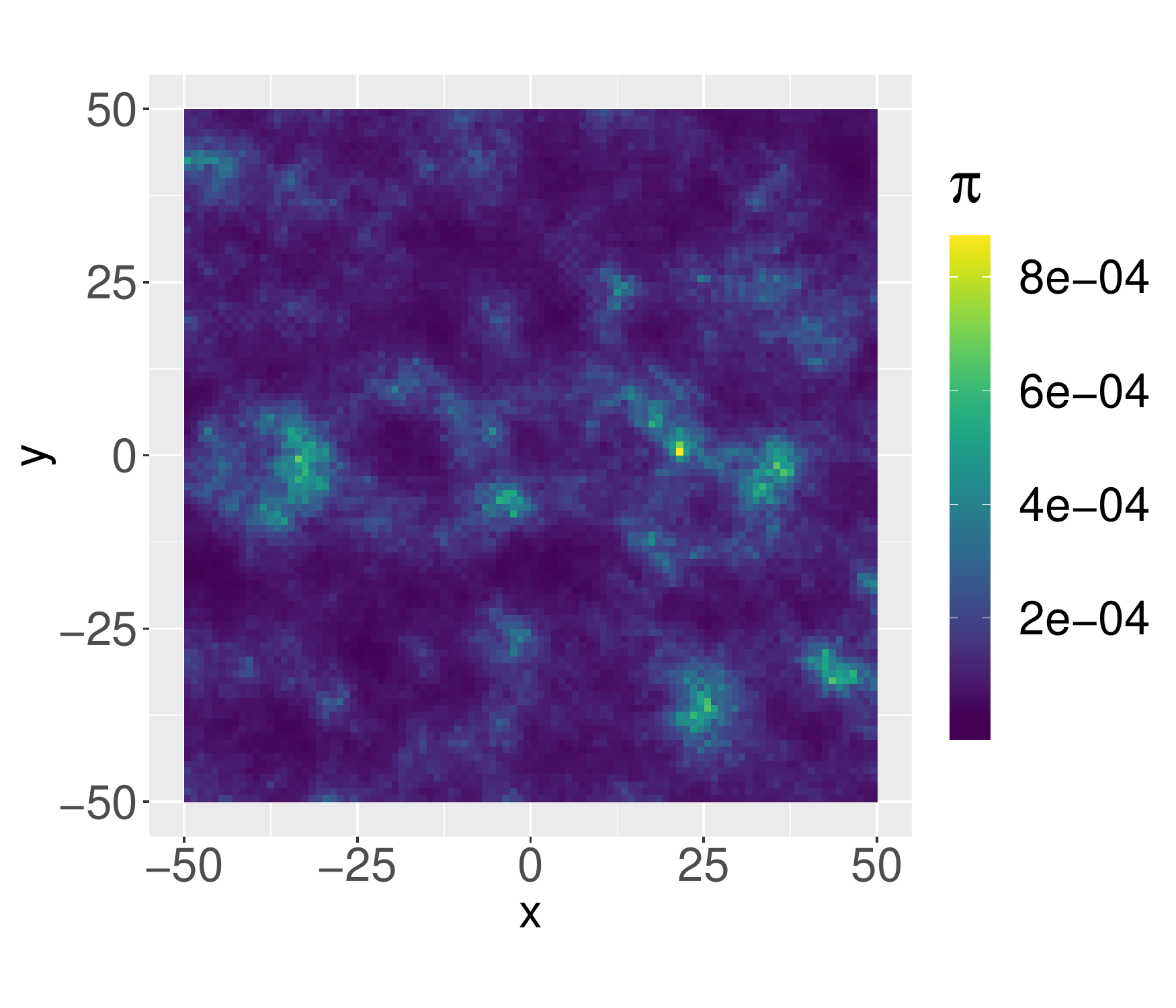}
    \caption{Simulated covariate fields $c_1$ and $c_2$, and utilisation distribution obtained with $\bm\beta = (2,4)'$. used in the second simulation scenario.}
    \label{fig:sim2_covs}
\end{figure}

To simulate covariates, we used a procedure similar to that described by \cite{avgar2016}. We defined a spatial grid over $[-50,50] \times [-50,50]$, with cells of size 1. 
For each cell, a random uniform value was generated on $(0,1)$, and the covariate field was obtained with a two-dimensional moving average filter over a circular region of radius $\rho$. Here, $\rho$ measures the degree of spatial autocorrelation of the simulated covariate. We then normalized the covariate field, to range between $0$ and $1$. 
Using this procedure, we simulated two covariates $c_1$ and $c_2$, with the same autocorrelation parameter $\rho=10$.
Then, we simulated trajectories from the Euler scheme described in Equation \eqref{eq:eulerApprox}, with $\pi$ defined as the (normalized) RSF with coefficients $(\beta_1, \beta_2) = (2,4)'$. 
Plots of the simulated covariates, and of the utilisation distribution used in the simulations, are shown in Figure \ref{fig:sim2_covs}.

We simulated 200 trajectories from the Langevin movement model, at a fine temporal resolution of $\Delta=0.01$. We then subsampled each trajectory, for different time resolutions $\Delta \in \{ 0.01, 0.02, 0.05, 0.1, 0.25, 0.5, 1 \}$, to emulate data sets obtained at different observation rates. From each thinned data set, we  kept the first 250 locations of each of the 200 trajectories, leading to a total of 50,000 locations.

\begin{figure}[htbp]
	\centering
    \includegraphics[width=0.32\textwidth]{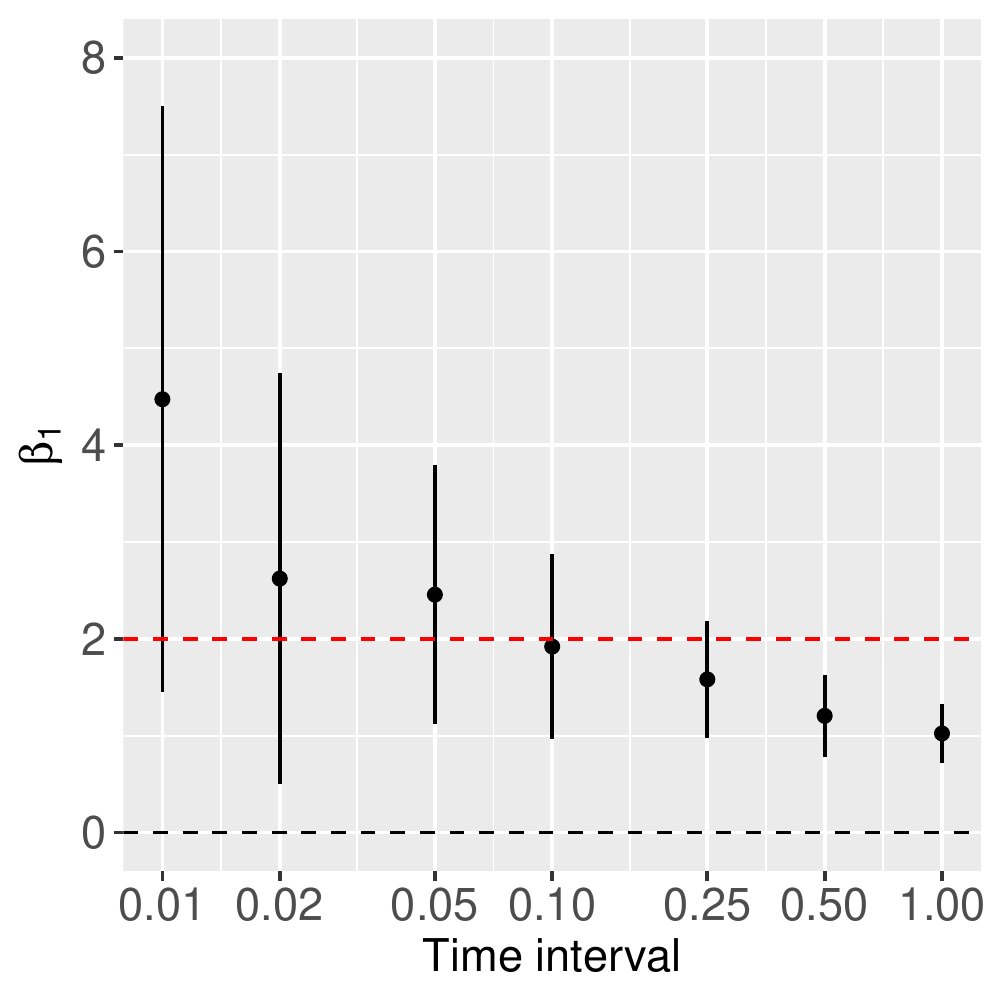}
    \includegraphics[width=0.32\textwidth]{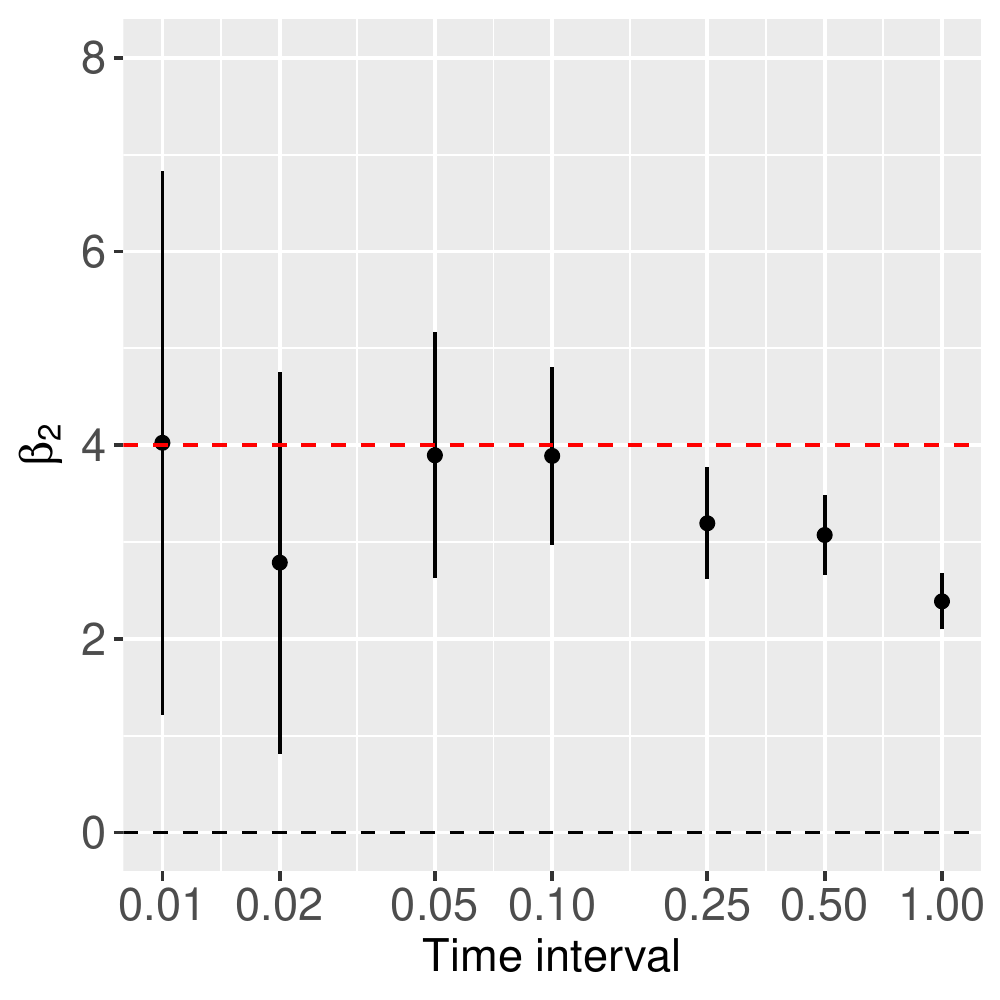}
    \includegraphics[width=0.32\textwidth]{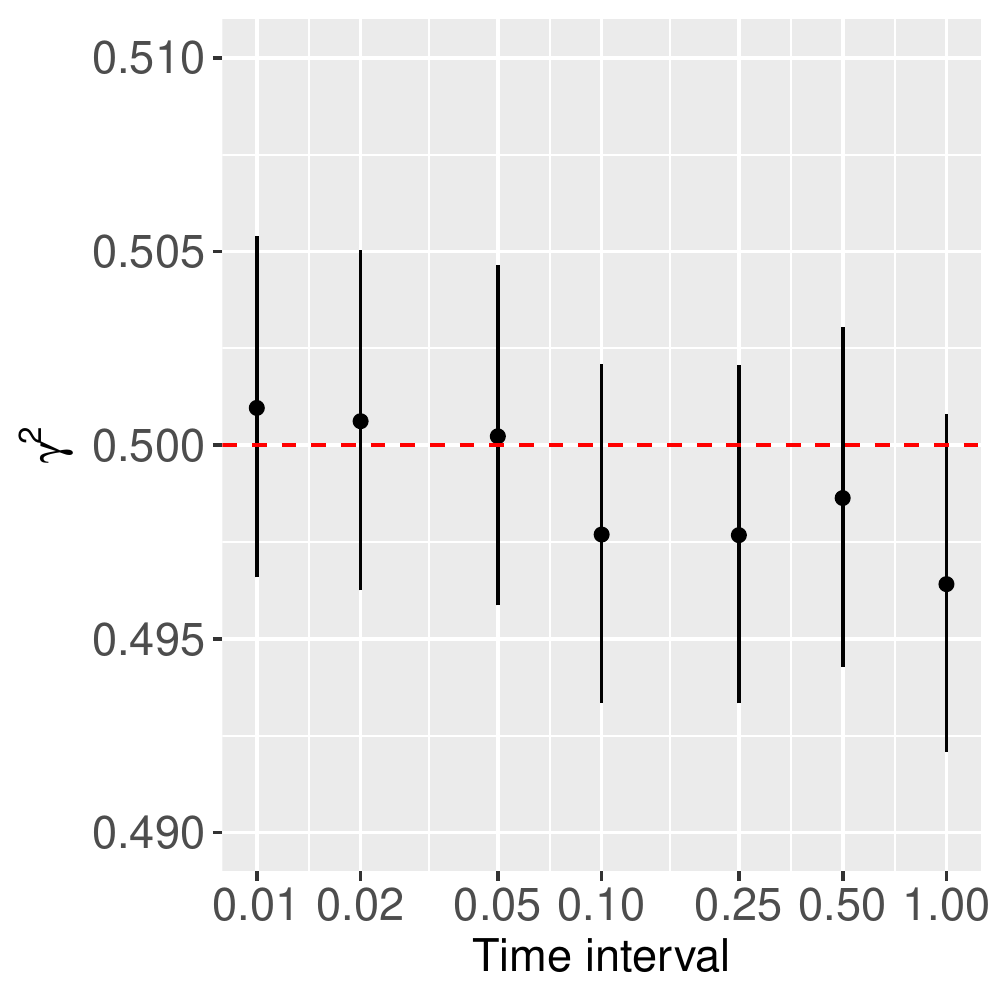}
    \caption{Estimates of the habitat selection parameters $\bm\beta$ and of the speed parameter $\gamma^2$ for different observation intervals, in the simulation study of Section \ref{sec:sim2}. The vertical lines show 95\% confidence intervals, and the red dotted lines are the true values of the parameters. The x axis is on the log scale.}
    \label{fig:sim2res1}
\end{figure}

We fitted the Langevin movement model to each thinned data set, using the estimators given in Section \ref{sec:MLE}. 
We evaluated the gradients of the covariates at each simulated location with a bilinear interpolation method. This procedure was done using the function \texttt{interp.surface} from the R package fields \citep{nychka2017}. 
We then obtained the approximated gradient of the interpolated covariates with the R package nloptr. 
Point estimates and 95\% confidence intervals of the habitat selection parameters $\bm\beta$ and the speed parameter $\gamma^2$ are displayed in Figure \ref{fig:sim2res1}.

The estimates of the speed parameter $\gamma^2$ were very close to the true value, for all simulation experiments. However, there was a lot of variability in the accuracy and precision of habitat selection parameter estimates. The uncertainty on the estimates of the habitat selection parameters decreased as the time interval increased. This is not surprising: all trajectories had the same number of locations, such that those with longer time intervals explored a larger proportion of the study region. Tracks with longer time intervals therefore covered a larger range of covariate values. Similarly to standard linear model analyses, the uncertainty on the coefficients is larger when the observed range of explanatory variables in Equation \eqref{eq:lin_model} is narrow. From $\Delta=0.05$ to $\Delta=1$, the estimates of $\beta_1$ and $\beta_2$ both decreased as the time interval increases, leading to an underestimation of the parameters for longer time intervals. This is a common problem for the estimation of discretely observed diffusion processes, because the consistency of the estimators requires $\Delta$ to tend towards 0 \citep[for more details, see][]{kessler2012statistical}. For long time intervals, the habitat selection parameters are underestimated in absolute value, i.e.\ the strength of the (positive or negative) effect is underestimated. This bias is a side effect of the underestimation of the speed of the process for long time lags between observations. As the time interval increases, the estimated utilisation distribution becomes flatter, to reflect our growing uncertainty about the effect of the covariates on the short-term movement. In the extreme, for very long time intervals, we would have no information about the selection process, and the estimated utilisation distribution would be flat, corresponding to a uniform distribution of space use over the study region. Note that, although the strength of selection was underestimated in the simulations with long time intervals, the sign of the effect -- i.e.\ selection or avoidance -- was always estimated correctly (Figure \ref{fig:sim2res1}).

To investigate the performance of the method for the analysis of data sets collected at irregular time intervals, we ran a similar experiment where the observations were thinned at random. 
The results were very similar to the simulations with regular intervals, and are presented in Appendix \ref{append:irregular}. 
These findings confirm that, due to its continuous-time formulation, the Langevin movement model can directly be used on tracking data collected irregularly.

\section{Example analysis}
\label{sec:casestudy}

\begin{figure}[htbp]
	\centering
    \includegraphics[width=\textwidth]{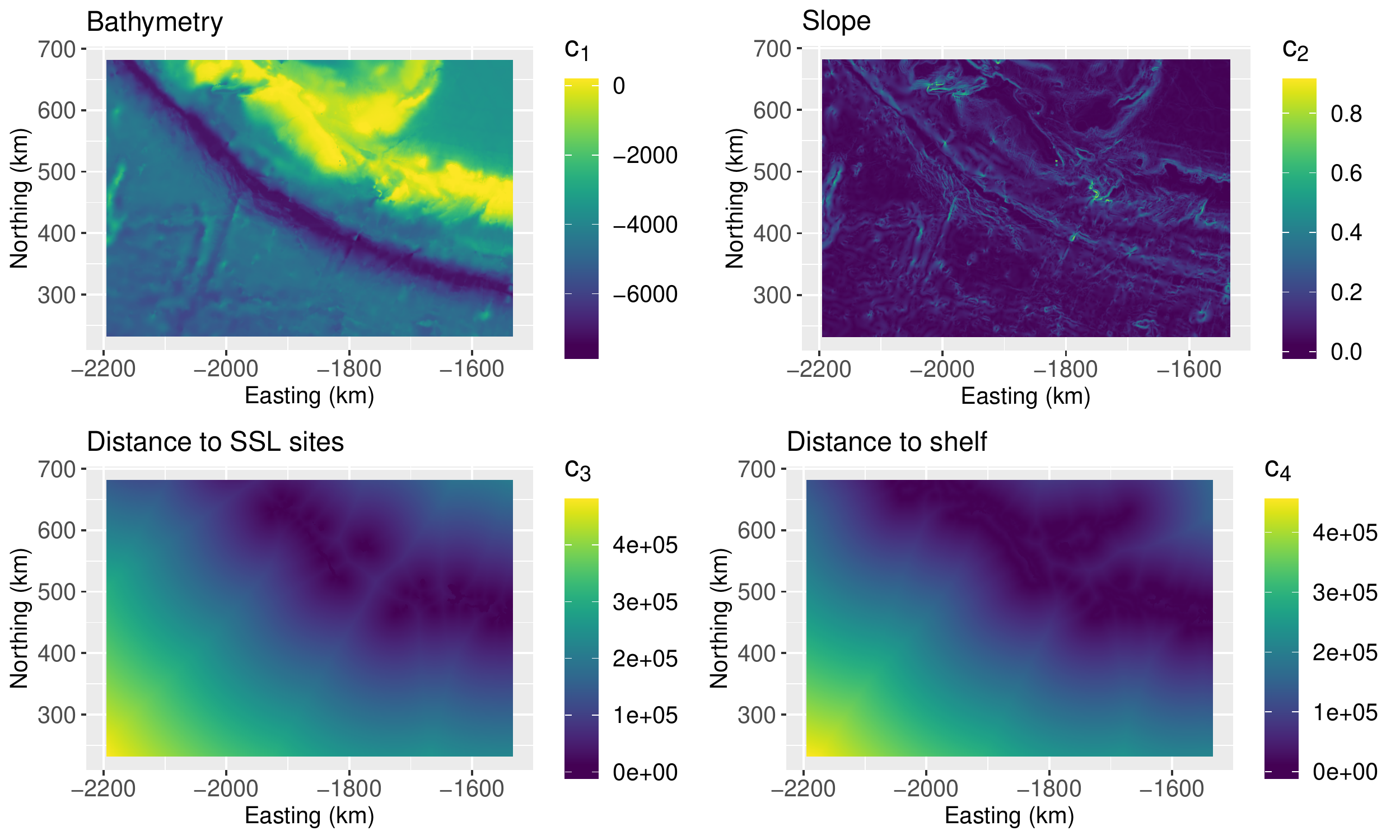}
    \caption{Covariate maps for the sea lion analysis.}
    \label{fig:SSLcovs}
\end{figure}

In this section, we fit the Langevin movement model to a data set described by \cite{wilson2018estimating}, collected on Steller sea lions (\emph{Eumetopias jubatus}) in Alaska. 
The data set comprises three trajectories, obtained from three different individuals, for a total of 2672 Argos locations. 
The time intervals were highly irregular, with percentiles $P_{0.025} = 6$min, $P_{0.5} = 1.28$h, $P_{0.975} = 17.4$h. 
In addition to the locations, \cite{wilson2018estimating} provided four spatial covariates over the study region, at a resolution of 1km: bathymetry ($c_1$), slope ($c_2$), distance to sites of interest ($c_3$), and distance to continental shelf ($c_4$). 
The sites of interest were either haul-out or rookery sites. 
Maps of the covariates are shown in Figure \ref{fig:SSLcovs}, and we refer the readers to \cite{wilson2018estimating} for more detail about the data set.

To correct for the measurement error, and to follow the preprocessing performed by \cite{wilson2018estimating}, we first fitted a continuous-time correlated random walk (CTCRW) to the tracks, using the R package crawl \citep{johnson2008, johnson2018}.
The CTCRW is a continuous-time state-space model, that can be used on irregular and noisy telemetry data. The package crawl implements the Kalman filter for this state-space model. 
We used the code provided by \cite{wilson2018estimating} to fit the CTCRW to each track, and obtained predicted locations for the times of the observations.

\begin{figure}[htbp]
	\centering
    \includegraphics[width=0.49\textwidth]{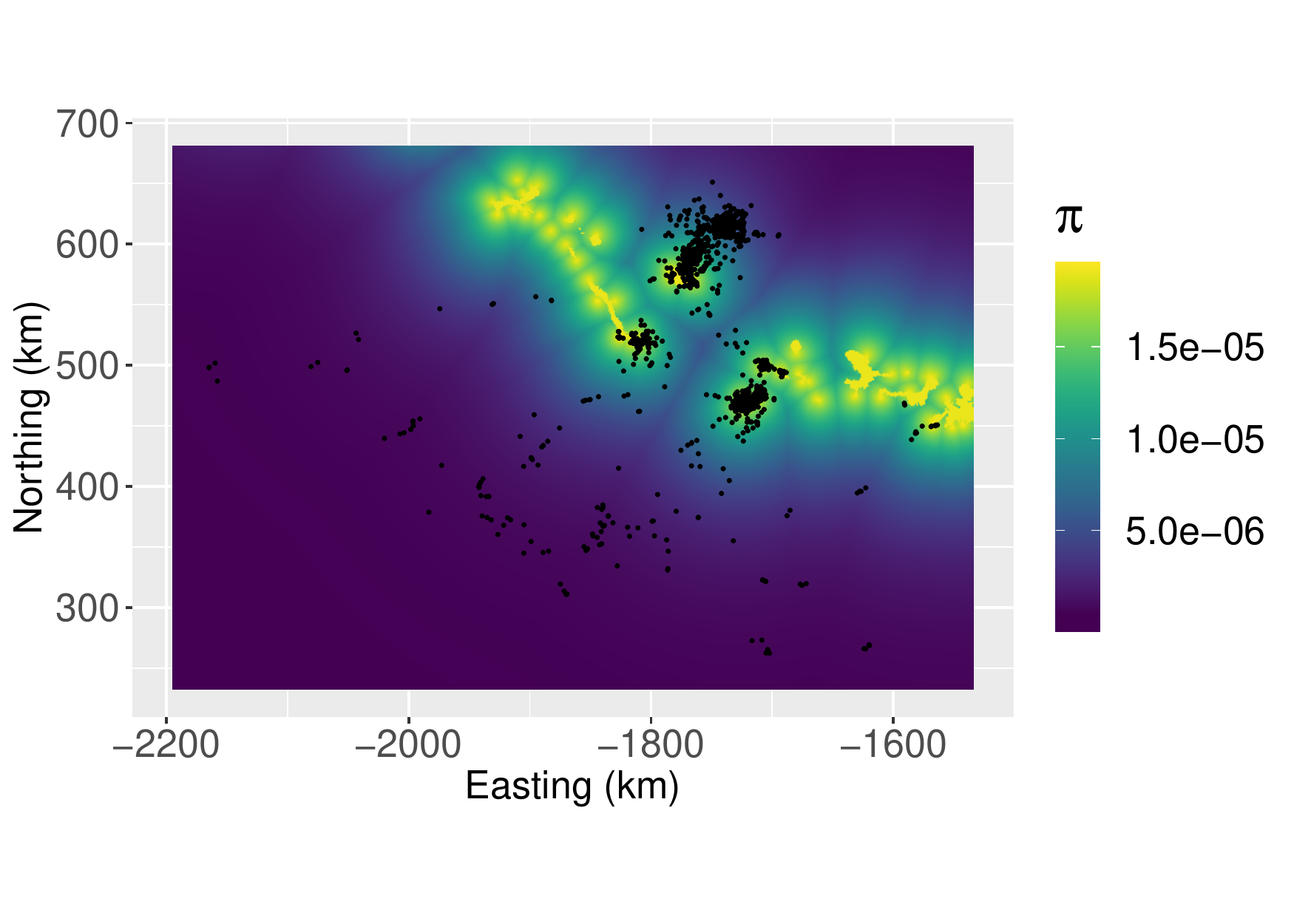}
	\includegraphics[width=0.49\textwidth]{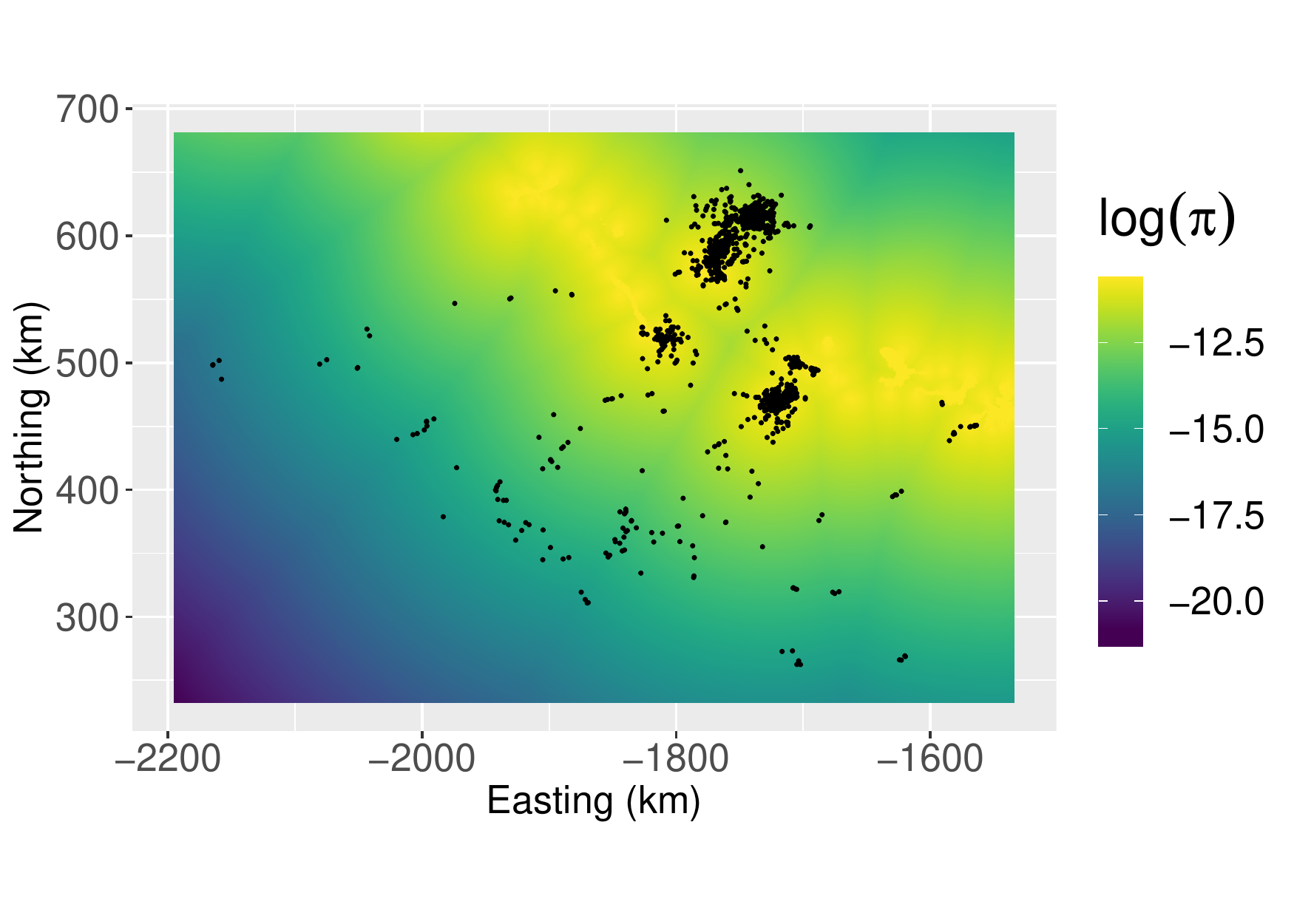}    
    \caption{Estimated utilisation distribution for the sea lion analysis (left), and its logarithm, for comparison with \cite{wilson2018estimating} (right). The black dots are the filtered sea lion locations.}
    \label{fig:SSLud}
\end{figure}

We then fitted the Langevin movement model to the filtered tracks, using the inference method of Section \ref{sec:inference}. 
Most of the computation time is needed to evaluate the gradient of each covariate at all observed locations, which took about 1.5 min on a 2GHz i5 CPU. 
Like in the simulation study of Section \ref{sec:sim2}, the covariates were interpolated, so that their gradient could be evaluated at each filtered location. 
The point estimates and 95\% confidence intervals, obtained from the equations of Section \ref{sec:MLE}, are given in Table \ref{tab:SSL}. The estimated utilisation distribution, and its logarithm \citep[for comparison with][]{wilson2018estimating}, are plotted in Figure \ref{fig:SSLud}. 

\begin{table}[htbp]
	\centering
    \begin{tabular}{ccc}
    	\toprule
    	& Estimate & 95\% CI \\
        \midrule
    	$\beta_1$ & $1.39 \cdot 10^{-4}$ & ($-3.87 \cdot 10^{-7}$, $2.79 \cdot 10^{-4}$) \\
 		$\beta_2$ & $0.12$ & ($-0.14$, $0.37$) \\
		$\beta_3$ & $-2.50 \cdot 10^{-5}$ & ($-3.58 \cdot 10^{-5}$, $-1.41 \cdot 10^{-5}$) \\
		$\beta_4$ & $3.47 \cdot 10^{-6}$ & ($2.05 \cdot 10^{-7}$, $6.73 \cdot 10^{-6}$) \\
        \bottomrule
    \end{tabular}
    \caption{Maximum likelihood estimates and 95\% confidence intervals obtained with the Euler scheme for the sea lion analysis.}
    \label{tab:SSL}
\end{table}

The 95\% confidence intervals of the parameters for two of the covariates (bathymetry and slope) include zero, i.e.\ we cannot draw conclusions about their effects on the sea lions' movement. 
The effect of the distance to sites of interest, $\beta_3$, was estimated to be negative. 
This indicates that the model captures the attraction of the sea lions towards the sites of interest (rookeries and haul-out sites). 
On the other hand, the effect of the distance to the shelf, $\beta_4$, was estimated to be positive. 
Although this seems to indicate that the animals tend to move away from the shelf, it may also be an artifact caused by the strong collinearity between the covariates $c_3$ and $c_4$.

We presented this analysis as a proof of concept for the Langevin movement model, and several directions could be explored further. As for any regression model, standard model selection criteria such as the AIC could be used to choose the best set of covariates. Following \cite{wilson2018estimating}, we could also estimate the parameters separately for the three seals, to capture inter-individual heterogeneity.

\section{Discussion}
\label{sec:Discussion}
This work introduces a new model of animal movement, based on the Langevin diffusion process, that integrates the movement with space use and habitat selection.
The movement model follows the idea of potential-based movement models proposed by \cite{preisler2004} and \cite{brillinger2010}, and it is explicitly connected to the animal's utilisation distribution, from stationarity properties of the Langevin diffusion process \citep{roberts1996}.
If spatial covariates are available, the long-term (utilisation) distribution can be modelled with a resource selection function, embedded in the movement process, to infer habitat preferences. 
The Langevin movement model therefore describes animal movement in response to spatial covariates, i.e.\ step selection. 
Pseudo-likelihood methods can be used to obtain estimates of the habitat selection parameters in a classical linear model framework, from which an estimated utilisation distribution can be computed. 
The Langevin movement model is formulated in continuous time, and it can deal with location data collected at irregular time intervals, without the need to interpolate them. Similarly, because it models movement in continuous space (unlike the method presented by \citealt{wilson2018estimating}), the interpretation of the results is not tied to a particular space discretisation.

In this paper, we used the Euler discretization scheme to approximate the likelihood of the model. This scheme is the most widely-used method to carry out inference for discretely-observed diffusion processes, when the transition density is not analytically tractable (see \citealp{preisler2004, brillinger2010, russell2018}, for applications in ecology).
There exist other pseudo-likelihood approaches, and \cite{gloaguen2018} argued that better inferences could be obtained with more refined schemes. 
In particular, they found that the Ozaki discretization provided more reliable results in their applications. 
However, the Ozaki scheme requires the evaluation of the partial derivatives of the drift, i.e.\ the (partial) second derivatives of $\log \pi$ in the Langevin movement model.
If $\pi$ is modelled with a resource selection function, then this would require the evaluation of the second derivatives of the covariate fields, from Equation \eqref{eq:gradlogpi}.
In practice, the covariates must be interpolated to a spatially continuous function, and their second derivatives computed using numerical methods. To compare the Euler and the Ozaki scheme, we repeated the simulation study of Section \ref{sec:sim1}, using the Ozaki scheme for the estimation. (The results are not shown here.) The theoretical advantages of the Ozaki scheme were counterbalanced by the need of this second-order interpolation, and the Euler scheme provided more reliable estimates. Therefore, in the context of the Langevin movement model, the Euler scheme is typically more robust to numerical approximations. 
To avoid any discretization scheme, it would be interesting to consider exact algorithms for diffusion processes, as suggested in \citep{gloaguen2018}. 
However, so far, exact methods of inference have been restricted to a small class of diffusions, and could not be readily applied to the cases presented here.

In the case study of Section \ref{sec:casestudy}, we used a two-stage approach to deal with the measurement error. 
We first fitted a state-space model, the continuous-time correlated random walk, to filter the Argos locations. 
Then, we fitted the Langevin movement model to the filtered tracks. 
There are several drawbacks to the two-stage approach. Indeed, it is difficult to propagate the uncertainty from the measurement error \citep[although multiple imputation could be used; see e.g.][]{scharf2017}. Besides, the two stages are not consistent, because the first stage ignores the environmental effects that are estimated in the second stage.
To avoid this issue, the two steps could be integrated into a state-space model that incorporates measurement error directly on top of the Langevin movement process. 
The state equation of the full model is given by the transition density of the Langevin movement model, or a discretization of it (like the one given in Equation \eqref{eq:eulerApprox}). 
A natural choice for the observation equation would be $\tilde{\bm{x}}_i = \bm{x}_i + \bm\eta_i$, where $\tilde{\bm{x}}_i$ is the noisy observed location, $\bm{x}_i$ is the true location, and $\bm\eta_i \sim N(\bm{0}, \sigma_\text{obs}^2 \bm{I}_2)$ models the measurement error. 
Under the Euler scheme, the approximate transition density is normal, and a Kalman filter can be used to compute the pseudo-likelihood of this hierarchical state-space model.
This extension is conceptually straightforward, and should be done in the near future.

\subsection*{Acknowledgements}
TM was supported by the Centre for Advanced Biological Modelling at the University of Sheffield, funded by the Leverhulme Trust, award number DS-2014-081. We thank Paul Blackwell for discussions during the early developments of the method.


\subsection*{Data accessibility}
The Steller sea lion data set used in Section \ref{sec:casestudy} is provided by \cite{wilson2018estimating}.

\bibliographystyle{apalike}
\bibliography{refs.bib}

\setcounter{table}{0}
\renewcommand{\thetable}{A\arabic{table}}
\newpage
\appendix

\section{Estimators}
\label{append:beta}

This appendix proves that $\hb$ and $\hat\gamma^2$ are unbiased estimators and derived the confidence intervals for the corresponding parameters.
 
\subsection{Unbiased estimators}

We defined $\hnu$ and $\hg^2$ as
\begin{align*}
	\hnu &= \left (\bm{D}^{\intercal} \bm{T}_{\Delta}^2  \bm{D} \right)^{-1} 
    	\bm{D}^{\intercal} \bm{T}_{\Delta} \bm{Y},\\
	\hg^2 &= \dfrac{1}{2n-J} \lVert \bm{Y} - \hat{\bm{Y}} \rVert^2.
\end{align*}

Following standard linear model properties, $\hnu \sim \mathcal{N}\left(\gamma^2\bm\beta, \left (\bm{D}^{\intercal} \bm{T}_{\Delta}^2 \bm{D} \right)^{-1} \right)$. Besides, Cochran's theorem implies that $\hnu$ and $\hg^2$ are independent and that $(2n - J)\hg^2/\gamma^2 \sim \chi^2(2n - J)$, where $\hg^2$ is an unbiased estimator of $\gamma^2$. 

Considering $\tilde{\bm\beta}$ defined as $\tilde{\bm\beta} = \hnu/\hg^2$, we have
\begin{align*}
	\mathbb{E}[\tilde{\bm\beta}] & = \mathbb{E}[\hnu]\mathbb{E}\left[\frac{1}{\hg^2}\right] \\ 
    & = \gamma^2\bm\beta \mathbb{E} \left[ \frac{1}{\hg^2} \right]\\
	& =  \gamma^2\bm\beta \times \frac{2n - J}{\gamma^2} \mathbb{E}\left[{\frac{1}{\frac{2n - J}{\gamma^2}\hg^2}}\right]
\end{align*}
From the properties given above, we have
\begin{equation*}
	\dfrac{1}{\frac{2n-J}{\gamma^2} \hg^2} \sim  \text{Inv-}\chi^2(2n - J),
\end{equation*}
and we obtain
\begin{equation*}
	\mathbb{E}[\tilde{\bm\beta}] = \frac{2n - J}{2n - J - 2} \bm\beta.
\end{equation*}

Thus, to obtain an unbiased estimator of $\bm\beta$, we define $\hb := (2n - J - 2)/(2n - J) \tilde{\bm\beta}$.

\subsection{Confidence intervals}

As said above,  $(2n - J)\hg^2/\gamma^2 \sim \chi^2(2n - J),$ so that a confidence interval for $\gamma^2$ is given by 

$$ CI_{\alpha}(\gamma) = \left [\hg\frac{ 2n-J }{ q_{{1-\alpha/2}, {2n - J}}} ; \ \hg\frac{ 2n-J }{ q_{{\alpha/2}, {2n - J}}}  \right ],
$$

where  $q_{{\alpha/2}, {2n - J}}$ stands for the quantile of order $\alpha/2$ from a $\chi^2$ distribution with $2n-J$ degrees of freedom.

It is also straightforward to derive a confidence interval or even a confidence ellipsoid using the distribution of $\hb$. 
We   first use the covariance matrix of $\tilde{\beta}, $
\begin{align*}
	\Cov(\tilde\beta_j, \tilde\beta_k) & = \mathbb{E} \left[ \frac{\hat\nu_j\hat\nu_k}{\hg^4} \right] - 
    	\mathbb{E} \left[ \frac{\hat\nu_j}{\hg^2} \right]
        \mathbb{E} \left[ \frac{\hat\nu_k}{\hg}^2 \right]
        & \text{by definition} \\
	& = \mathbb{E}[\hat\nu_j\hat\nu_k] \mathbb{E} \left[ \frac{1}{\hg^4} \right] - 
    	\mathbb{E} \left[ \frac{1}{\hg^2} \right]^2 \mathbb{E}[\hat\nu_j]\mathbb{E}[\hat\nu_k] 
        & \text{by independence of the estimators $\hg$ and $\hnu$.}
\end{align*}
We now note that
\begin{align*}
	\mathbb{E}[\hat\nu_j\hat\nu_k] & = \Cov(\hat\nu_j,\hat\nu_k) + \mathbb{E}[\hat\nu_j]\mathbb{E}[\hat\nu_k] \\
    & = \gamma^2\Upsilon_{jk} + \beta_j\beta_k\gamma^4,
\end{align*}
where $\Upsilon_{jk} := \left[ (\bm{D}^{\intercal} \bm{T}_{\Delta}^2  \bm{D} )^{-1} \right]_{jk}$. Moreover:
\begin{align*}    
	\mathbb{V}\left[\frac{1}{\hg^2}\right] 
    	& =  \frac{2(2n - J)^2}{\gamma^4(2n - J - 2)^2(2n - J - 4)} 
        & \text{as } \frac{\gamma^2}{(2n - J)\hg^2} \sim \text{Inv-}\chi^2(2n - J)\\
	\mathbb{E}\left[\frac{1}{\hg^4}\right] & = 
    	\mathbb{V} \left[ \frac{1}{\hg^2} \right] + \mathbb{E} \left[ \frac{1}{\hg^2} \right]^2 = 
        \frac{(2n - J)^2}{\gamma^4(2n - J - 2)^2}\left(\frac{2}{2n - J - 4} + 1\right).&
	\intertext{Finally, we can rewrite}
	\Cov(\tilde\beta_j, \tilde\beta_k) & = \mathbb{V}\left[\frac{1}{\hg^2}\right] 
    	\mathbb{E}[\hat\nu_j]\mathbb{E}[\hat\nu_k] + \mathbb{E}\left[\frac{1}{
        \hg^4}\right]\Cov(\hat\nu_j,\hat\nu_k) &\\
	& = \frac{(2n - J)^2}{(2n - J - 2)^2} \left\lbrace \frac{2\beta_j\beta_k}{2n - J - 4} + 
    	\frac{\Upsilon_{jk}}{\gamma^2} \left(1 + \frac{2}{2n - J - 4} \right) \right\rbrace.
\end{align*}

\section{Simulation parameters}
\label{append:simpar}

The simulation parameters used in Section \ref{sec:sim1} are given in Table \ref{tab:simpar}.

\begin{table}[htbp]
	\centering
    \begin{tabular}{ccc}
    	\toprule
    	& $j=1$ & $j=2$ \\
    	\midrule
    	$\alpha_j$ & $6$ & $6$\\
    	$a_1^j$ & $0$ & $-2$ \\
    	$a_2^j$ & $0$ & $\frac{\pi}{2}$ \\
    	$\omega_1^j$ & $0.6$ & $0.1$ \\
    	$\omega_2^j$ & $0.2$ & $0.5$ \\
    	$\sigma_1^j$ & $0.4$ & $0.4$ \\
   	 	$\sigma_2^j$ & $0.4$ & $0.4$ \\
    	\bottomrule
    \end{tabular}
    \caption{Simulation parameters for Section \ref{sec:sim1}.}
	\label{tab:simpar}
\end{table}

\section{Simulation study with irregular intervals}
\label{append:irregular}

We repeated the second simulation scenario (Section \ref{sec:sim2}), with irregular time intervals. Starting from the same full simulated data set ($\Delta=0.01$), we thinned the observations at random, to obtain irregularly-sampled locations. We ran two experiments: (1) mean time interval $\bar\Delta = 0.05$, (2) mean time interval $\bar\Delta = 0.50$. The estimates and 95\% confidence intervals of the habitat selection parameters $\beta_1$ and $\beta_2$ are given in Table \ref{tab:irregular}. We compare them to the results of the simulations conducted in Section \ref{sec:sim2} with regular intervals. The estimates obtained in the simulations with regular and irregular time intervals are very similar, and the standard errors are virtually identical.


\begin{table}[htbp]
	\centering
	\begin{tabular}{ccccc}
		\toprule
        & \multicolumn{2}{c}{regular} & \multicolumn{2}{c}{irregular} \\
        \midrule
        Mean interval & $\bar\Delta = 0.05$ & $\bar\Delta = 0.5$ & $\bar\Delta = 0.05$ (SD: $0.045$) & $\bar\Delta = 0.5$ (SD: $0.49$) \\
        \midrule
    	$\beta_1$ & $2.46$ (SE: $0.68$) & $1.20$ (SE: $0.22$) & $2.15$ (SE: $0.69$) & $0.89$ (SE: $0.22$) \\
        $\beta_2$ & $3.89$ (SE: $0.65$) & $3.07$ (SE: $0.21$) & $3.84$ (SE: $0.65$) & $2.90$ (SE: $0.21$) \\
        \bottomrule
	\end{tabular}
    \caption{Comparison of estimates and standard errors of the habitat selection parameters, for regular and irregular time intervals of simulation.}
    \label{tab:irregular}
\end{table}

This confirms that the continuous-time formulation of the Langevin movement model can accommodate irregular time intervals, without the need to interpolate prior to the analysis.

\end{document}